\documentclass[11pt,a4paper,twoside,groupcitations]{article}
\usepackage[]{latexsym,amsmath,amssymb}
\usepackage[]{epsfig}
\usepackage{color}
\usepackage{graphicx}
\usepackage{cite}
\usepackage{stackrel}
\usepackage{amsfonts, bbold, mathtools, physics, cancel, scalerel, stackengine, wasysym, subcaption}
\newcommand{\be}{\begin{eqnarray}}
\newcommand{\ee}{\end{eqnarray}}
\def\beq{\begin{equation}}
\def\eeq{\end{equation}}

\newcommand{\Det}{\text{Det}\,}
\title{\bf Massive graviton from diffeomorphism invariance}
\author{
João~M.~L.~de~Freitas$^{a}$\thanks{E-mail: matheus.leal@ufpr.br},
$\ $
Iber\^e Kuntz$^{a}$\thanks{E-mail: kuntz@fisica.ufpr.br}
\\
\\
$^a${\em Departamento de F\'isica, Universidade Federal do Paran\'a}
\\
{\em PO Box 19044, Curitiba -- PR, 81531-980, Brazil}
\\
\\
}
\date{}
\begin{document}
\maketitle
\begin{abstract}
We describe a mechanism in which the graviton acquires a mass from the functional measure without violating the diffeomorphism symmetry nor including St\"uckelberg fields. Since gauge invariance is not violated, the number of degrees of freedom goes as in general relativity. For the same reason, Boulware-Deser ghosts and the vDVZ disconinuity do not show up. The graviton thus becomes massive at the quantum level while avoiding the usual issues of massive gravity. 
\end{abstract}
%
%\textit{PACS 04.60 - Quantum theory of gravitation.}
%\end{abstract}
%
%
%\pacs{}
%
%
%%%%%%%%%%%%%%%%%%%%%%%%%%%%%%%%%%%%%%%%%%%%%%%%%%%%%%%%%%%%%%%%
%%%
%%%                     INTRODUCTION
%%%
%%%%%%%%%%%%%%%%%%%%%%%%%%%%%%%%%%%%%%%%%%%%%%%%%%%%%%%%%%%%%%%%
%
\newpage
\section{Introduction}
\label{S:intro}
\setcounter{equation}{0}

General relativity and the standard model have both been extremely successful in describing fundamental phenomena. Nonetheless, many problems in the interface of gravity and high-energy physics, such as the dark sector and singularities, cannot be explained by either of these theories. This has led to a plethora of attempts to modify these models.

One interesting modification of general relativity regards massive gravity. Attempts of giving the graviton a mass dates back to the 30s \cite{Fierz:1939ix}, but it was only recently that consistent theories of massive gravity have been found \cite{Dvali:2000xg,Dvali:2000hr,Dvali:2000rv,Bergshoeff:2009hq,deRham:2010kj} (see also Ref.~\cite{deRham:2014zqa} for an in-depth review). A massive graviton indeed introduces many issues, such as the violation of diffeomorphism (gauge) invariance, the van Dam-Veltman-Zakharov (vDVZ) discontinuity and the presence of Boulware-Deser ghosts \cite{Boulware:1974sr}. Gauge invariance can be reinstated via St\"uckelberg fields, which is legitimate but does require new degrees of freedom. The vDVZ discontinuity \cite{vanDam:1970vg,Zakharov:1970cc}, namely the disagreement with general relativity in the massless limit, is usually conjectured to be solved at the non-linear regime by the Vainshtein screening mechanism \cite{Vainshtein:1972sx,Babichev:2013usa}. Only a few models have successfully implemented the Vainshtein mechanism \cite{Babichev:2013usa,Babichev:2009jt,Deffayet:2001uk,Nicolis:2008in,deRham:2016plk}, which can also bring along the possibility of superluminal velocities \cite{Dvali:2000hr,Nicolis:2008in,Luty:2003vm,Nicolis:2004qq,Adams:2006sv,deFromont:2013iwa}.

In this paper, we propose a novel procedure to give the graviton a mass without introducing any of the aforementioned issues and without modifying classical general relativity. Our finding is based on a non-trivial path-integral measure, which is required for obtaining gauge invariant correlation functions. The functional measure introduces non-linear loop corrections, which act as a gravitational potential and result in a (quantum) mass for the graviton in the linear regime. The most important point is the preservation of the diffeomorphism invariance, which is responsible for keeping the theory free of ghosts and of the vDVZ discontinuity. The so obtained mass is, however, pure imaginary, thus precluding the existence of gravitons as asymptotic states.

This paper is organized as follows. In Sec.~\ref{measure1}, we review some aspects of the functional measure in quantum field theory. The one-loop correction induced by the functional measure is then studied in Sec.~\ref{massg}, where the graviton mass is calculated. Comparison with experimental data yields stringent bounds on the model. In Sec.~\ref{newton}, we obtain Newton's potential and discuss its consequences. Finally, we draw our conclusions in Sec.~\ref{conc}.

\section{Functional measure in quantum field theory}
\label{measure1}

A central object in quantum field theory is the generating functional:
\begin{equation}
	Z[J]
	=
	\int\mathrm{d}\mu[\varphi] e^{i \left( S[\varphi^i] + J_i \varphi^i \right)},
	\label{Z}
\end{equation}
where $S[\varphi]$ is the classical action for some generic set of fields $\varphi^i = (\phi(x), A_\mu(x), g_{\mu\nu}(x), \ldots)$. All information regarding any physical system is contained in its corresponding generating functional $Z[J]$. In particular, correlation functions are obtained by simple functional differentiations, from which scattering amplitudes can be obtained using the LSZ (Lehmann--Symanzik--Zimmermann) formula. Despite its importance in field theory, a rigorous mathematical foundation remains unknown, particularly with respect to the functional measure $\mathrm{d}\mu[\varphi]$. Operationally, one can however define the aforementioned object as \cite{Mottola:1995sj,DeWitt:2003pm,Toms:1986sh,Casadio:2022ozp}
\begin{equation}
    \mathrm{d}\mu[\varphi] = \mathcal{D}\varphi^i \sqrt{\Det G_{ij}}
    \ ,
    \label{measure}
\end{equation}
where $\mathcal{D}\varphi^i = \prod_i \mathrm{d}\varphi^i$ and $\Det G_{ij}$ denotes the functional determinant of the configuration-space metric $G_{ij}$. The factor $\sqrt{\Det G_{ij}}$ is required to account for a non-trivial configuration space, whose typical example is that of a non-linear sigma model.

The presence of a non-trivial functional measure is usually sidestepped by writing $\delta^{(4)}(0) = 0$ in dimensional regularization, in which case one finds
\begin{align}
    \Det G_{ij}
    &=
    e^{\delta^{(4)}(0) \int\mathrm{d}^4x \sqrt{-g} \, \tr\log G_{ij}}
    \nonumber
    \\
    &\stackrel{dim.\, reg.}{=} \ 1
    \ .\label{dimr}
\end{align}
One should, however, be extremelly careful when setting such extreme infinities to zero. In fact, if such regularization were always correct, the Jacobian would always be unit and anomalies, for example, would never take place. 
Rather than hiding the problem under the umbrela of formal manipulations, it is safer to implement lattice or cutoff regularizations when defining the path integral. This becomes particularly important in Wilson's effective field theory, where a natural cutoff is implemented. 

In the Wilsonian spirit, we shall adopt Gaussian regularization
\begin{equation}
    \delta^{(4)}(x)
    =
    \frac{\Lambda^4}{(2\pi)^{2}} e^{\frac{-x^2 \Lambda^2}{2}}
    \ ,
\end{equation}
for some (soft) cutoff $\Lambda$. For $\delta^{(4)}(0) = \tfrac{\Lambda^4}{(2\pi)^{2}}$, we find \cite{Kuntz:2022kcw}:
\begin{equation}
	Z[J]
	=
	\int \mathcal{D}\varphi^i e^{i \left( S_\text{eff}[\varphi^i] + J_i \varphi^i \right)},
\end{equation}
with the Wilsonian effective action
\begin{equation}
    S_\text{eff} = \int\mathrm{d}^4x \sqrt{-g}
    \left(
    \mathcal L
	-
    i \zeta \, \tr\log G_{IJ}
	\right)
	\ ,
	\label{genact}
\end{equation}
for some bare Lagrangian $\mathcal L$. Here $\zeta=\zeta(\Lambda)$ is a Wilsonian coefficient whose running is such that
\begin{equation}
    \Lambda \frac{dZ[J]}{d\Lambda} = 0
    \label{RGE}
    \ .
\end{equation}
Note that the Dirac delta divergence is polynomial. The coefficient $\zeta$ is thus expected to be UV sensitive.

One should note that the configuration-space metric $G_{ij}$ must be seen as part of the definition of the theory. Although it is typically identified with the bilinear form in the kinetic term \cite{Meetz:1969as,Slavnov:1971mz,Vilkovisky:1984st,Fradkin:1973wke,Fradkin:1976xa}, this identification is not based on physical reasonings. Nonetheless, the functional measure must be invariant under the underlying symmetries, which allows one to determine $G_{ij}$ in the same spirit as effective field theories. In particular, for pure gravity this procedure results in the well-known DeWitt metric \cite{DeWitt:1967yk}. To lowest order, the most general configuration-space metric for arbitrary fields yields \cite{Casadio:2021rwj,Kuntz:2022tat,Casadio:2020zmn}

\begin{equation}
    S_\text{eff} = \int\mathrm{d}^4x \sqrt{-g}
    \left(
    \mathcal L
	-
    i \gamma \, \tr\log |g_{\mu\nu}|
	\right)
	\ ,
	\label{eq:newac}
\end{equation}
where $\gamma$ is obtained from $\zeta$ via a finite renormalization.

In spite of the form of the correction, we stress that Eq.~\eqref{eq:newac} does not violate diffeomorphism invariance. The apparent violation results from the fact that $\sqrt{\Det G_{ij}}$ transforms as a (functional) scalar density, thus so does the last term in Eq.~\eqref{eq:newac}. However, the $\mathcal{D}\varphi^i$ also transforms as a scalar density in such a way that the full measure $\mathcal{D}\varphi^i \sqrt{\Det G_{ij}}$ is invariant. Therefore, variations of the apparent symmetry-breaking term under spacetime diffeomorphisms are canceled by the functional Jacobian that shows up from $\mathcal{D}\varphi^i$, keeping the quantum theory and all observables invariant~\footnote{Strictly speaking, the invariance of off-shell quantities, such as the effective action, requires a connection in configuration space \cite{Vilkovisky:1984st,DeWitt:1988dq}. Because we are mainly interested in the phenomenology of the functional measure, we shall not dwell on this topic.}.
Because the functional measure and the classical action are invariant under diffeomorphisms, the background-field effective action $\Gamma[g]$ naturally reflects such symmetry. In fact, at the one-loop level it reads
\begin{align}
	\Gamma[\varphi^i]
	&=
	S[\varphi^i]
	- \frac{i}{2} \log\Det G_{ij}
	+ \frac{i}{2} \log\Det \mathcal{H}_{ij}
	\label{inv1loop}
	\\
	&=
	S[\varphi^i]
	+ \frac{i}{2} \log\Det \mathcal{H}^i_{\ j}
	\ ,
\end{align}
where the configuration-space metric $G_{ij}$ enters the usual correction $\log\Det\mathcal{H}_{ij}$, transforming the bilinear map $\mathcal{H}_{ij} = S_{,ij}$, whose determinant is basis-dependent, into a linear operator $\mathcal{H}^i_{\ j}$, whose determinant is invariant \cite{Toms:1986sh,Ellicott:1987ir}.

Finally, the above findings were obtained in the Lorentzian path integral. The imaginary factor in Eq.~\eqref{genact} (hence in Eq.~\eqref{eq:newac}) shows up when the $i=\sqrt{-1}$ in the argument of the exponential in the Lorentzian path integral is pulled out to write the measure as a correction to the classical action. Defining the functional measure in the Euclidean formalism
\begin{equation}
	Z_E[J]
	=
	\int\mathrm{d}\mu[\varphi] e^{- \left( S_\text{eff}^E[\varphi^i] + J_i \varphi^i \right)}
\end{equation}
yields a real one-loop correction:
\begin{equation}
	S_\text{eff}^E
	=
	\int\mathrm{d}^4x \sqrt{-g}
    \left(
    \mathcal L
	-
    \gamma \, \tr\log |g_{\mu\nu}|
	\right)
	\ .
	\label{acE}
\end{equation}
One thus faces the problem of whether the path-integral measure should be defined in the Euclidean space (and rotated back to real time) or straight in the Lorentzian space. The former is required for a better mathematical construction of the path integral, albeit still largely formal. On the other hand, as far as our current experiments are concerned, Nature is fundamentally Lorentzian. For this reason, we shall perform our calculations by defining the measure on the latter. Naturally, predictions shall be different in different schemes. At this level of formality, only time will tell which one, if any, is correct.

\section{Pure gravity as an example: the DeWitt metric}
\label{pure}

As a concrete example, let us consider the Einstein-Hilbert action:
\begin{equation}
	S = \int\mathrm{d}^4x \sqrt{-g} \frac{M_p^2}{2} R
	\ ,
	\label{eq:EH}
\end{equation} 
where $M_p$ is the reduced Planck mass and $R$ the Ricci scalar.
In this case, the Hessian reads:
\begin{equation}
	\mathcal{H}_{\mu\nu\rho\sigma}
	=
	K_{\mu\nu\rho\sigma}
	\Box
	+ U_{\mu\nu\rho\sigma}
	\ ,
	\label{hessianGR}
\end{equation}
where
\begin{equation}
	K_{\mu\nu\rho\sigma} = \frac14
	\left(
		g_{\mu\rho} g_{\nu\sigma}
		+ g_{\mu\sigma} g_{\nu\rho}
		- g_{\mu\nu} g_{\rho\sigma}
	\right)
\end{equation}
and $U_{\mu\nu\rho\sigma}$ is a tensor that depends on the spacetime curvature. The precise form of $U_{\mu\nu\rho\sigma}$ is unimportant to us and can be found, for example, in Ref.~\cite{Percacci:2017fkn}.
The simplest configuration-space metric for pure gravity is given by the so-called DeWitt metric:
\begin{equation}
	G_{\mu\nu\rho\sigma}
	=
	\frac12
	(g_{\mu\rho} g_{\nu\sigma}
	+ g_{\mu\sigma} g_{\nu\rho}
	- a \, g_{\mu\nu} g_{\rho\sigma})
	\ ,
	\label{DWmetric}
\end{equation}
which depends on a dimensionless parameter $a$.
Considering Eqs.~\eqref{hessianGR} and \eqref{DWmetric}, the combination \eqref{inv1loop} results in:
\begin{align}
	\Gamma[g]
	=
	\int\mathrm{d}^4x \sqrt{-g}
	\Bigg\{&
		\frac{M_p^2}{2} R
		+ \frac{i\zeta}{2}
		\log\det
		\left[
		\frac12
		\left(
			\delta_{( \mu}^{\ \ \rho} \delta_{\nu )}^{\ \ \sigma}
			+ (a-1) g_{\mu\nu}g^{\rho\sigma}
		\right)
		\right]
		\nonumber
	\\
	&+ \frac{i}{2}
	\log\det \left[\delta_{( \alpha}^{\ \ \mu} \delta_{\beta )}^{\ \ \nu}\Box + (K^{-1})^{\mu\nu\rho\sigma}U_{\rho\sigma\alpha\beta}\right]
	\Bigg\}
	\ ,
	\label{1piac}
\end{align}
where we used the lowercase $\det$ to denote the ordinary (finite-dimensional) determinant.
Note that the indices turn out to be at the correct position, having the same number of covariant and contravariant indices. Under diffeomorphisms, the determinant will always produce equal factors of the Jacobian and its inverse, canceling them out and leaving the effective action invariant (as one expects from the correct transformation of the functional measure \eqref{measure}).
This is, in fact, the reason one can generate a mass for the graviton without violating the gauge symmetry.

The last term in Eq.~\eqref{1piac} can be computed by employing asymptotic expansions either in the curvature or spacetime derivatives \cite{DeWitt:2003pm,Barvinsky:1987uw,Barvinsky:1985an}. As such, at low energies, they are subdominant in comparison with the second term, which contains no factor of curvature or derivative and corresponds to the functional measure contribution. We can thus focus on the first line of Eq.~\eqref{1piac}. 
The matrix determinant lemma can be used to write:
\begin{equation}
	\det\left[\delta^I_{\ J} + (a-1) g_J g^I \right]
	=
	1 + 4(a-1)
	\ ,
\end{equation}
thus Eq.~\eqref{1piac} can be massaged into:
\begin{equation}
	\Gamma[g]
	=
	\int\mathrm{d}^4x \sqrt{-g}
	\left[
		\frac{M_p^2}{2} R
		+ i\Lambda_{C}
	\right]
	\ ,
	\label{1piac2}
\end{equation}
where we defined
\begin{equation}
	\Lambda_C = 
		\frac{\zeta}{2}
		\log
		\left[
		\frac{
				1
				+ 4(a-1)
		}{256}
		\right]
		\ .
\end{equation}
For the DeWitt metric, we conclude that the functional measure gives a complex contribution to the cosmological constant.

Performing a metric perturbation around Minkowski
\begin{equation}
	g_{\mu\nu}
	=
	\eta_{\mu\nu}
	+ \frac{2}{M_p} 
	h_{\mu\nu}
	\label{pert}
\end{equation}
in Eq.~\eqref{1piac2} leads to
\begin{align}
	\Gamma =
	\int \mathrm{d}^4x
	\bigg[
		&
		i\frac{\Lambda_C}{M_p} h
		- \frac{1}{2} \partial_{\lambda} h_{\mu \nu} \partial^{\lambda} h^{\mu \nu}
		+ \frac{1}{2} \partial_{\lambda} h \partial^{\lambda} h
		- \partial_{\mu} h^{\mu \nu} \partial_{\nu} h
		+ \partial_{\mu} h_{\nu \lambda} \partial^{\nu} h^{\mu \lambda}
		\nonumber
		\\
		&
		+ \frac{i}{2} \frac{\Lambda_C}{M_p^2} (h^2 - 2 h_{\mu \nu} h^{\mu \nu})
	\bigg]
	\ .
	\label{1piac3}
\end{align}
One should note the appearance of a non-vanishing linear term in the graviton field, which corresponds to a tadpole. This term results from the expansion around a background that is not a solution to the effective field equations. The presence of a cosmological constant indeed prevents the Minkowski background from being the vacuum solution. Nonetheless, since the cosmological constant only showed up as a quantum correction, we stress that Eq.~\eqref{pert} is perfectly legitimate as the leading contribution in perturbation theory. At the one-loop level, evaluating correlation functions at $\eta_{\mu\nu}$ indeed only produces errors at $\mathcal{O}(\hbar^2)$. Therefore, as it is customary in quantum field theory, one-loop correlation functions can be evaluated at tree-level solutions.

The tadpole is what usually tells apart the cosmological constant from the graviton mass. Interpreting the cosmological constant as a mass would require the removal of the tadpole. In scalar field theories, tadpoles are easily removed by shifting the scalar field by a constant. However, such a field redefinition in the gravitational context, namely 
\begin{equation}
	h_{\mu\nu} \to h_{\mu\nu} - \frac{M_p}{2}\eta_{\mu\nu}
	\ ,
\end{equation}
would cancel out the background metric in Eq.~\eqref{pert}. Instead, the tadpole can be removed by
\begin{equation}
	h_{\mu\nu} \to h_{\mu\nu} + h^{(0)}_{\mu\nu}
	\ ,
\end{equation}
where $h^{(0)}_{\mu\nu}(x)$ is a non-dynamical spacetime-dependent field chosen to cancel the tadpole. Alternatively, the gravitational tadpole can be canceled by a cosmological counter-term, chosen so that the linear term in Eq.~\eqref{1piac3} is exactly zero \cite{Capper:1973bk,deWit:1977av}. Either way, the result would amount on a finite renormalization of $\Lambda_C$, thus one can simply drop the tadpole. We can, therefore, identify:
\begin{equation}
	m^2
	=
	i\frac{\zeta}{2 M_p^2}
		\log
		\left[
		\frac{
				1
				+ 4(a-1)
		}{256}
		\right]
\end{equation}
as the graviton mass.

At last, one might worry that a ghost is present due to the non-Fierz-Pauli combination of the mass terms in Eq.~\eqref{1piac3} . We stress, however, that these terms only showed up at the quantum level. The particle spectra, on the other hand, is obtained from the classical action Eq.~\eqref{eq:EH}, which is pure general relativity, thus containing only 2 degrees of freedom. These are the only degrees of freedom that become massive.

\section{Massive graviton}
\label{massg}
In the last section, we used a particular choice for the configuration-space metric. Although DeWitt's metric is quite simple and already capable of generating a mass, it is far from unique. When other types of fields are present in addition to the spacetime metric, the configuration-space metric can get very difficult to handle \cite{Casadio:2021rwj}. To leading order in the fields, however, we obtained Eq.~\eqref{eq:newac} as the general functional measure correction for arbitrary fields in curved spacetime. In light of such correction, we now generalize the results of Sec.~\ref{pure}. 

When the bare Lagrangian $\mathcal L$ is the Einstein-Hilbert term, the functional measure changes the dynamics of space-time as follows~\footnote{We adopt the metric signature $(-+++)$.}:
\begin{equation}
    S_\text{eff} = \int\mathrm{d}^4x \sqrt{-g}
    \left(
    \frac{M_p^2}{2} R
	-
    i \gamma \, \tr\log |g_{\mu\nu}|
    + \mathcal{L}_m
	\right)
	\ ,
	\label{effac}
\end{equation}
where $\mathcal{L}_m$ is the Lagrangian for matter fields.
The corresponding equations of motion read~\footnote{One should note that $T_{\mu\nu}$ is covariantly conserved in the sense of the Slavnov-Taylor identities, i.e. $\langle \nabla^\mu T_{\mu\nu} \rangle = 0$. In particular, such conservation holds at every loop order.}:
\begin{equation}
	G_{\mu\nu} + i \frac{2\gamma}{M_p^2} \left[1 + \frac12 \log(-g)\right] g_{\mu\nu} = \frac{1}{M_p^2} T_{\mu\nu}
	\ ,
	\label{eom}
\end{equation}
where $T_{\mu\nu}$ is the energy-momentum tensor for $\mathcal{L}_m$. One should note that general relativity is smoothly recovered in the limit $\gamma\to 0$. The parameter $\gamma$ is proportional to the graviton mass (see Eq.~\eqref{mass}), thus there is no tension with the experimental tests of general relativity. As we shall see in the next section, gauge invariance prevents the vDVZ discontinuity from appearing.
  
Using Eq.~\eqref{pert} in Eq.~\eqref{effac} gives:
\begin{align}
	S_\text{eff} =
	\int \mathrm{d}^4x
	\bigg[
		&
		- \frac{1}{2} \partial_{\lambda} h_{\mu \nu} \partial^{\lambda} h^{\mu \nu}
		+ \frac{1}{2} \partial_{\lambda} h \partial^{\lambda} h
		- \partial_{\mu} h^{\mu \nu} \partial_{\nu} h
		+ \partial_{\mu} h_{\nu \lambda} \partial^{\nu} h^{\mu \lambda}
		\nonumber
		\\
		&
		- \frac{i}{2} m^2_g (h^2 - h_{\mu \nu} h^{\mu \nu})
		+ M_p^{-1} h_{\mu\nu} T^{\mu\nu}
	\bigg]
	\ ,
	\label{eq:FP}
\end{align}
where we have already dropped the tadpole by the arguments outlined in the end of the last section~\footnote{Because the relative coefficient in the mass terms in Eq.~\eqref{eq:FP} is different from Eq.~\eqref{1piac3}, in this case the tadpole can also be removed by the constant shift $h_{\mu\nu}\to  h_{\mu\nu} -\frac{M_p}{6}\eta_{\mu\nu}$ without canceling out the background metric.}.
The graviton mass is thus given by
\begin{equation}
	m^2 \equiv i m^2_g = \frac{4 i \gamma}{M_p^2}
	\ .	
	\label{mass}
\end{equation}
Notice that the mass term in Eq.~\eqref{eq:FP} comes with the correct relative coefficient between $h^2$ and $h_{\mu\nu}h^{\mu\nu}$ for a ghost-free theory. We stress that such a coefficient is not finely tuned by hand, it follows directly from the quantum correction due to the functional measure.
Comparing the modulus of Eq.~\eqref{mass} to the bound found by LIGO \cite{LIGOScientific:2016lio}:
\begin{equation}
	m_g < 1.2 \times 10^{-22} \, \text{eV}
\end{equation}
translates into a bound on $\gamma$:
\begin{equation}
	\gamma < 2 \times 10^{-26} \, \text{GeV}^4 \ . 
\end{equation}

A few comments are in order. Because of \eqref{RGE}, the graviton mass \eqref{mass} runs with the energy scale.
%\begin{equation}
%	\Lambda\frac{\d m^2}{\d\Lambda}
%	=
%	- \frac{i\beta}{2\pi^2 M_p^2} \Lambda^4
%%	- \frac{i n \beta}{16\pi^{2} M_p \sqrt{\gamma(\Lambda)}} \Lambda^n
%	\ .
%\end{equation}
In particular, in the classical limit $\hbar \to 0$, the functional measure correction vanishes and so does the graviton mass $m$. We stress that the massless limit $m\to 0$ (or, equivalently, $\gamma\to 0$) is smooth as can be seen from the non-linear theory Eq.~\eqref{eom}.
%{\color{red} Therefore, no vDVZ discontinuity or strong-coupling issue arises for $m\to 0$. Vainshtein screening mechanism is thus not needed, which avoids issues with superluminal velocities.}
Therefore, all tests of general relativity are automatically satisfied. Secondly, since the gauge symmetry is not broken, the counting of degrees of freedom goes as usual for general relativity. Namely, a symmetric second rank tensor contains 10 degrees of freedom, 8 of which can be eliminated by gauge transformations, yielding only 2 propagating modes rather than 5. This follows because one has not started with massive gravity \textit{ab initio}. Indeed, since the functional measure does not contain derivatives, no new degrees of freedom show up and the spectrum continues to be determined from the classical Einstein-Hilbert action. The mass is generated only after quantization for the propagating modes that had already been present in the classical theory.
As a result, Boulware-Deser ghosts do not appear in the theory as the action Eq.~\eqref{effac} does not contain higher derivatives.
Finally, we stress that our proposal has also the advantage of being a top-down approach. We indeed know the non-linear theory of massive gravity from the onset.

The resulting mass squared \eqref{mass} is, however, pure imaginary. The imaginary part of the mass is a measure of its lifetime. Indeed, the position-space propagator gains a decreasing time exponential:
\begin{equation}
	- i D_{\mu\nu\rho\sigma}
	\sim
	\int\frac{\mathrm{d}^3\vec p}{(2\pi)^3}
	\,
	\frac{e^{i \vec p \cdot (\vec x - \vec x')}}{2\omega}
	\left[
		\theta(t-t') e^{- i \omega (t-t')}
		+ \theta(t'-t) e^{- i \omega (t'-t)}
	\right]
\end{equation}
with
\begin{equation}
	\omega
	=
	\left(|\vec p|^4 + m_g^4\right)^{1/4}
	\left[
		\cos\frac{\theta}{2}
		-i \sin\frac{\theta}{2}
	\right]
	\label{eq:freq}
\end{equation}
and $\theta = \arctan(m_g^2/|\vec p|^2)$. Notice that the imaginary part of the frequency \eqref{eq:freq} approaches zero as $m_g\to 0$, in which case $\omega \to |\vec p|$.
Such imaginary part yields a damping behaviour, killing off the graviton's perturbations at large times.
The recent detection of gravitational waves thus put stringent bounds on $m_g$.

\section{Newtonian potential}
\label{newton}

Gravitons with complex mass also affect the mediation of the gravitational force.
An immediate consequence of such a massive graviton regards the modification on the Newtonian potential. As a result of the presence of mass, one usually expects a Yukawa potential. However, because the graviton mass is complex, the resulting potential shall develop an oscillating behavior modulated by the Yukawa decay, as we shall now see.

From Eq.~\eqref{eq:FP}, we obtain the effective equations of motion for the graviton:
\begin{equation}
    \Box h_{\mu \nu}
    - \Box h \eta_{\mu \nu}
    + \partial_{\mu} \partial_{\nu} h
    + \partial_{\alpha} \partial_{\beta} h^{\alpha \beta} \eta_{\mu \nu}
    - \partial^{\lambda} \partial_{\mu} h_{\lambda \nu} - \partial^{\lambda} \partial_{\nu} h_{\lambda \mu}
    =
    - M_p^{-1} T_{\mu \nu}
    - m^2 ( h_{\mu \nu} -  h \eta_{\mu \nu})
    \label{eomugly}
\end{equation}
The divergence and the trace of Eq.~\eqref{eomugly} read
\begin{align}
	\label{div}
    \partial^{\mu} h_{\mu \nu}
    &=
	\partial_\nu h
	\\
	\label{trace}
	h
	&=
	\frac{M_p^{-1}}{3 m^2} T
	\ .
\end{align}
Using Eqs.~\eqref{div}--\eqref{trace} in Eq.~\eqref{eomugly} gives
\begin{equation}
	\label{conservedgrav}
    (\Box + m^2) h_{\mu \nu}
    =
    - M_p^{-1} \left(
    	T_{\mu \nu}
    	- \frac{1}{2} T \eta_{\mu \nu}
    \right)
    + \left(
    	\partial_\mu\partial_\nu h
    	- \frac12 \eta_{\mu\nu} m^2 h
    \right)
    \ .
\end{equation}
The first term on the RHS of Eq.~\eqref{conservedgrav} is the usual general relativistic result (apart from the mass term on the LHS), which is then modified by the second term on the RHS, leading to the vDVZ discontinuity. In our case, because gauge invariance is not broken, the second term can be eliminated by a choice of gauge. In fact, under a diffeomorphism such a term transforms as
\begin{equation}
	\partial_\mu\partial_\nu h - \frac12 \eta_{\mu\nu} m^2 h
    \to
    \partial_\mu\partial_\nu h - \frac12 \eta_{\mu\nu} m^2 h
    + 2 \partial_\mu\partial_\nu\partial_\alpha \xi^\alpha
    - \eta_{\mu\nu} m^2 \partial_\alpha \xi^\alpha
    \ .
\end{equation}
Thus choosing $\partial_\alpha \xi^\alpha = -h/2$ results in:
\begin{equation}
	(\Box + m^2) h_{\mu \nu}
    =
    - M_p^{-1} \left(
    	T_{\mu \nu}
    	- \frac{1}{2} T \eta_{\mu \nu}
    \right)
    \ .
    \label{massiveh}
\end{equation}
One should notice the apperance of the factor of $1/2$ instead of the infamous $1/3$ of gauge-violating massive theories. Eq.~\eqref{massiveh} shows that the mass produced by the functional measure is perfectly consistent with all general relativistic tests as no vDVZ discontinuity takes place.

One can easily solve Eq.~\eqref{massiveh} in momentum space:
\begin{equation}
	\label{gensolcon}
    h_{\mu \nu}
    =
    M_p^{-1}
    \int \frac{\dd[4]{p}}{(2\pi)^4}
    \frac{e^{i p_{\alpha} x^{\alpha}}}{p^2 - m ^2}
    \left[
    	\tilde{T}_{\mu \nu}
    	- \frac{1}{2} \eta_{\mu \nu} \tilde{T}
    \right]
    \ .
\end{equation}
For a static point source of mass $M$ at the origin:
\begin{equation}
    T_{\mu \nu} = M \delta_{\mu 0} \delta_{0 \nu} \delta(\vec{x})
    \ ,
\end{equation}
one finds
\begin{equation}
    h_{0 0}
    =
    \frac12 \frac{M}{M_p} \frac{1}{4 \pi r} e^{ i m r}
    \ .
    \label{eq:newton}
\end{equation}
Despite Eq.~\eqref{eq:newton} having a Yukawa-like functional form, the complex exponential leads to novel predictions. Indeed, its real part provides the Newtonian potential~\footnote{We recall that $h_{00} = -2V$.}
\begin{align}
	V(r)
	=
	- \frac{M}{16 \pi M_p} \frac{e^{ - \tfrac{m_g}{\sqrt{2}}  r}}{r} \cos\left(\frac{m_g}{\sqrt{2}} \, r\right)
	\ ,
	\label{newton2}
\end{align}
whereas, by the optical theorem, its imaginary part relates to the total cross section~\footnote{Note that the imaginary part of Eq.~\eqref{eq:newton} remains finite at $r=0$ because $\sin(x)/x \to 1$.}. At small distances $r \ll \sqrt{2}/m_g$, our result recovers Newton's potential:
\begin{equation}
	V(r)
	=
	- \frac{M}{16 \pi M_p \, r}
	\left(1 - \frac{m_g}{\sqrt{2}} r + \frac{m_g^3}{6 \sqrt{2}} r^3 \right)
	+ \cdots
	\ .
\end{equation}
The leading correction is constant, thus does not affect the dynamics, so the first measurable new effect shows up only at next-to-leading order. 
We see that Newton's potential is modified at large distances $r \sim \sqrt{2}/m_g$.

\begin{figure}[htb]
	\begin{subfigure}{0.5\textwidth}
  		\centering
		\includegraphics[scale=0.5]{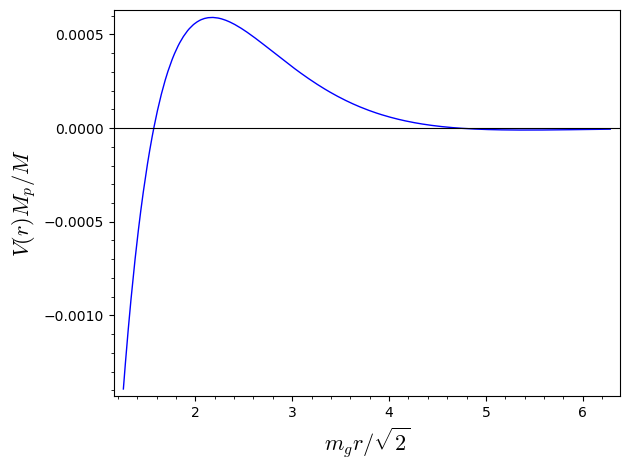}  
  		\caption{First and highest barrier.}
  		\label{fig:pot}
	\end{subfigure}
	\begin{subfigure}{0.5\textwidth}
		\centering
		\includegraphics[scale=0.5]{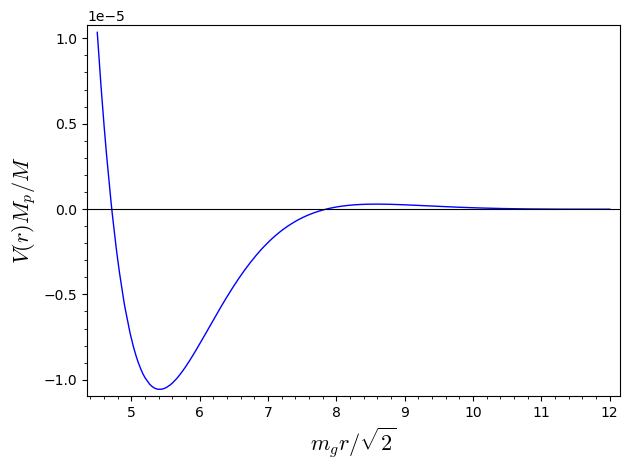}
		\caption{First and deepest potential well.}
		\label{fig:pot2}
	\end{subfigure}
	\caption{Plot of the Newtonian potential corrected by the functional measure, as given by Eq.~\eqref{newton2}.}
\end{figure}

The cosine function in Eq.~\eqref{newton2} can turn the potential's derivative positive, thus creating islands of bounded motion of decreasing depth. At each of these islands, gravity becomes repulsive. But because of the utterly small graviton mass, such an effect is only felt at very large distances:
\begin{equation}
	r > \frac{\pi\sqrt{2}}{2} m_g^{-1}
	\ .
\end{equation}
At the first and highest barrier $r_0 \sim 3.1 \, m_g^{-1}$, the potential height is given by (see Figure~\ref{fig:pot})
\begin{equation}
	V(r_0) \sim 4.2 \times 10^{-3} \, \frac{M m_g}{M_p}
	\ .
	\label{1well}
\end{equation}
If there is enough energy to overcome this barrier, the system alternates between regions of attractive and repulsive gravity as the distance increases. The system could get trapped between two consecutive barriers, thus forming a gravitational bound state, should its energy be smaller than the potential well (see Figure~\ref{fig:pot2}). This effect, however, is rapidly weakened by the exponential suppression in Eq.~\eqref{newton2}. Even at the deepest well depicted in Figure~\ref{fig:pot2}, the existing energy from surrounding astrophysical events is likely enough to keep such bound states from forming.
On the other hand, should the energy be smaller than Eq.~\eqref{1well}, the system would not collapse as the distance decreases, thus resolving the singularity at $r=0$. Note that the ratio $m_g/M_p$ is utterly negligible in Eq.~\eqref{1well}, thus only very massive objects, such as black holes, could prevent such collapse from happening.

\section{Conclusions}
\label{conc}

Understanding the quantum nature of gravity requires, among other things, grasping the graviton kinematics and dynamics. Quantum corrections trigger new graviton self-interactions, which affect the graviton dynamics. As in any gauge theory, it is generally believed that the diffeomorphism invariance precludes the appearance of mass terms in the effective action of general relativity. However, this follows by side-stepping the non-trivial Jacobian that shows up in the path integral when diffeomorphisms (or, more generally, field redefinitions) are performed. A proper regularization of the Jacobian shows that it cannot be simply ignored.

A diffeomorphism-invariant functional measure thus requires the factor $\sqrt{\Det G_{ij}}$ to cancel out the functional Jacobian, in very much the same way that $\sqrt{-g}$ is needed for integrations in curved spacetimes. This additional factor contributes as a quantum effective potential for the graviton, giving it a mass in the linear regime. Therefore, general covariance is actually the reason for the existence, rather than for the absence as commonly thought, of a mass for the graviton. 

We stress that such a mass is not obtained by modifying general relativity. One is simply quantizing general relativity by properly considering the geometry of configuration space, which reflects in the definition of the functional measure. One can view different configuration-space metrics as different quantization approaches to the same classical theory. The naive approach, where the functional Jacobian is disregarded, corresponds to a flat configuration space.
Unfortunately, there is no known physical principle other than symmetry to help us determining the configuration-space metric. On symmetry grounds, the lowest-order configuration-space metric is curved and depends on the spacetime metric. This situation is dramatically different for Yang-Mills fields, where the lowest-order configuration-space metric is trivial and no gauge-invariant higher-order term exists. Therefore, gauge invariance manifests itself quite differently in gravity than in the other interactions, being the sole responsible for keeping Yang-Mills fields massless (in the unbroken vacuum) and the graviton massive.
\section*{Acknowledgments}
IK is grateful to the National Council for Scientific and Technological Development -- CNPq (Grant No. 303283/2022-0) for partial financial support.

%%%%%%%%%%%%%%%%%%%%%%%%%%%%%%%%%%%%%%%%%%%%%%%%%%%%%%%%%%%%%%%%%
%%%
%%%                     BIBLIOGRAPHY
%%%
%%%%%%%%%%%%%%%%%%%%%%%%%%%%%%%%%%%%%%%%%%%%%%%%%%%%%%%%%%%%%%%%%
%
%
%
%

%
\end{document}